\documentclass[twocolumn,final,prd,superscriptaddress,showpacs,floatfix%
,preprintnumbers,nofootinbib]{revtex4-1}

\usepackage{verbatim}
\usepackage[final]{epsfig}
\usepackage{bm}
\usepackage[usenames]{color}

\newcommand{\I}{{\rm i}}
\newcommand{\E}{{\rm e}}

\begin{document}

\title{Fast time variations
of supernova neutrino signals from 3-dimensional models}

\author{Tina Lund}
\affiliation{Department of Physics, North Carolina State University,
2401 Stinson Drive, Raleigh, NC 27607, USA}

\author{Annop Wongwathanarat}
\affiliation{Max-Planck-Institut f\"ur Astrophysik,
Karl-Schwarzschild-Str.~1, 85748 Garching, Germany}

\author{Hans-Thomas Janka}
\affiliation{Max-Planck-Institut f\"ur Astrophysik,
Karl-Schwarzschild-Str.~1, 85748 Garching, Germany}

\author{Ewald M\"uller}
\affiliation{Max-Planck-Institut f\"ur Astrophysik,
Karl-Schwarzschild-Str.~1, 85748 Garching, Germany}

\author{Georg Raffelt}
\affiliation{Max-Planck-Institut f\"ur Physik (Werner-Heisenberg-Institut), 
F\"ohringer Ring 6, 80805 M\"unchen, Germany}

\date{31 July 2012}

\preprint{MPP-2012-122, INT-PUB-12-035}

\begin{abstract}
We study supernova neutrino flux variations in the IceCube detector,
using 3D models based on a simplified neutrino transport scheme. The
hemispherically integrated neutrino emission shows significantly
smaller variations compared with our previous study of 2D models,
largely because of the reduced SASI activity in this set of 3D
models which we interpret as a pessimistic extreme. For the studied
cases, intrinsic flux variations up to about 100~Hz frequencies
could still be detected in a supernova closer than about 2~kpc.
\end{abstract}


\maketitle

\section{Introduction}

Measuring the neutrino signal from a future galactic core-collapse
supernova (SN) will provide valuable information about the
conditions in the collapsing stellar core and in the forming compact
remnant. In particular neutrinos and gravitational waves will serve
as direct probes of the explosion mechanism. While neutrino energy
deposition is most widely favored as the trigger and energy supply
of the SN blast wave \cite{Bethe:1985}, the success of this
mechanism in iron-core SNe turns out to be tightly linked to the
development of violent nonradial mass flows in the layer behind the
stalled bounce shock (see Ref.~\cite{Janka:2012} for a recent
review).

Multi-dimensional hydrodynamical models have shown that nonradial
gas motions naturally grow from small initial perturbations in the
neutrino-heated, shocked accretion flow because of convective
instability~\cite{Bethe:1990} and the standing accretion shock
instability (SASI)~\cite{Blondin:2002sm}. Violent, quasi-periodic
expansion and contraction (``sloshing'') of the shock can lead to
considerable variation of the mass accretion rate to the
proto-neutron star (PNS). In axi-symmetric (2D) models, where the
sloshing motions are artificially constrained to proceed along the
symmetry axis, these variations and the associated compression of
the PNS ``surface'' layer can produce fluctuations of the observable
(hemispherically integrated) neutrino emission of more than 10\% in
luminosity and of 0.5--1\,MeV in the mean spectral energy; the polar
emission variations are even larger~\cite{Marek:2008qi}. Such large
variations of the neutrino emission would be easily detectable for a
galactic SN with IceCube or future megaton-class instruments, and
the presence of Fourier components with frequencies of tens of Hz up
to 200--300 Hz would provide important information on the SN core
dynamics prior to the onset of the explosion
\cite{Lund:2010,Brandt:2011}. They could also allow one to constrain
neutrino masses \cite{Ellis:2012a} and to probe neutrino propagation
over cosmic distances \cite{Ellis:2012b}.

Because of the directing effect of the symmetry axis the SASI
sloshing motions appear particularly strongly in 2D core-collapse
simulations, independently of whether detailed neutrino transport
\cite{Marek:2008qi, Marek:2007gr, Mueller:2012jm, Mueller:2012jh,
Brandt:2011, Suwa:2010}, simplified neutrino cooling and heating
terms \cite{Murphy:2008, Nordhaus:2010, Hanke:2011}, a schematic
cooling prescription, or an outflow boundary at the bottom of the
otherwise adiabatic accretion flow~\cite{Blondin:2002sm,
Blondin:2006} is used.

In three-dimensional (3D) models, low-order multipole SASI sloshing
motions have so far been observed only with considerably reduced
amplitudes and stochastically changing direction \cite{Iwakami:2008}
or no clear signatures of SASI sloshing activity were seen
\cite{Wongwathanarat:2010jm, Mueller:2012jw, Takiwaki:2012,
Burrows:2012}. Indeed, it has been claimed that buoyancy-driven
convection dominates post-shock turbulence \cite{Murphy:2012db} and
the SASI is ``at most a minor feature of SN dynamics''
\cite{Burrows:2012}. Such conclusions, however, should be taken with
caution. They rely on small sets of 3D models with idealized setups,
e.g., only including simple neutrino heating and cooling terms but
without detailed neutrino transport and its feedback on the PNS
evolution. They therefore do not explore the important influence of
PNS contraction driven by energy and lepton losses through neutrino
emission as well as progenitor-specific differences, both of which
were shown to be able to fundamentally change the dynamics of the
postshock accretion flow \cite{Marek:2007gr, Janka:2012,
Mueller:2012jh}. Moreover, the conditions for developing SASI spiral
modes \cite{Blondin:2006,Iwakami:2009,Fernandez:2010} in SN cores
are neither fully understood nor satisfactorily investigated. The
growth timescale of such modes can be strongly reduced by even small
amounts of rotation in the collapsing stellar matter
\cite{Yamasaki:2008} and their appearance is observed in
experimental analogies \cite{Foglizzo:2012}. Low-mode asymmetries,
which are much bigger than seen in present, first 3D simulations and
which impose quasi-periodic, large-amplitude variability on the PNS
neutrino emission, can therefore not be excluded.

We here take the pessimistic point of view and apply our previous
analysis of 2D models \cite{Lund:2010} to several recent 3D
explosion calculations which show only a low level of SASI activity
and no signs of long-lasting sloshing motions \cite{Mueller:2012jw},
in contrast to the 2D simulations of Marek, Janka and
M\"uller~\cite{Marek:2008qi} evaluated by Lund et
al.~\cite{Lund:2010}. Nevertheless, even for such unfavorable
conditions our results suggest that the modulation of the neutrino
emission below about 100~Hz will be detectable with IceCube for a
galactic SN up to a distance of 2~kpc.

The outline of our paper is as follows.
In Sec.~\ref{sec:numerical} we summarize essential information
about the simulated models and the expected signal rates in
IceCube. In Sect.~\ref{sec:power} we present the power spectra
that can be deduced from a measurement and discuss their
meaning. In Sect.~\ref{sec:shotnoise} we explore the
detectability of the intrinsic neutrino signal variations
in competition with the shot noise of the IceCube measurement.
Conclusions will be given in Sec.~\ref{sec:conclusion}.

\section{Supernova models}\label{sec:numerical}

\subsection{Investigated models}       \label{sec:models}

The 3D simulations providing the neutrino luminosities and mean
spectral energies for our analysis are models W15-4, L15-3, and
N20-2 from the larger set of explosion models published
in~\cite{Mueller:2012jw}, where also more details about the
numerical methods and setup can be found. These simulations are
based on three nonrotating progenitors, the 15\,$M_\odot$ model
s15s7b2 of Woosley and  Weaver \cite{Woosley:1995}, a 15\,$M_\odot$
star computed by Limongi et al.~\cite{Limongi:2000}, and a
20\,$M_\odot$ model provided by Shigeyama and
Nomoto~\cite{Shigeyama:1990}, respectively. In
Table~\ref{tab:datachart} some characteristic properties of our
three investigated explosion models are listed.

\begin{table}[b!]
\centering \caption{\label{tab:datachart} Properties of our models.}
\begin{tabular*}{\linewidth}{@{\extracolsep{\fill}}l r r r r}
\hline
\hline
Model   &               L15-3& W15-4 & N20-2    \\
\hline
Simulation end time [ms] &         1414 & 1315 & 1314   \\
Explosion time [ms] &   477     & 272   & 265   \\
Analysis interval length $\tau$ [ms] & 500     & 300   & 350   \\
$N_{\rm bins}$	& 500	& 300	& 350 \\
$N_{\rm events}$ [$10^7$] at 1 kpc	& 2.89	& 1.72	& 2.50 \\
Frequency resolution $\delta f$ [Hz] &  2.0 & 3.3 & 2.9  \\
$P_{0}$ [$10^9$] at 1~kpc & 3.34 & 3.29 & 5.10 \\
$P_{\rm comb}$ at 1~kpc & 346 & 573 & 612 \\
$P_{\rm bkgd}$ & 8.0 & 13.4 & 11.5 \\
$P_{99}$ at 2~kpc & 399 & 660 & 705 \\
\hline
\end{tabular*}
\end{table}

The 3D explosion modeling was performed with the
explicit, finite-volume, Eulerian, multi-fluid hydrodynamics
code {\sc Prometheus} \cite{Fryxell:1991,Mueller:1991a,Mueller:1991b}.
An axis-free, overlapping ``Yin-Yang'' grid technique in
spherical polar coordinates was used \cite{Kageyama:2004,Wongwathanarat:2010},
employing an angular resolution of 2$^\circ$ over the full 4$\pi$ solid angle.
A combined linearly and logarithmically spaced grid with 400 radial zones
allowed for a minimum resolution of 0.3\,km between the inner grid boundary
and the outer boundary at 18,000\,km. The high-density core of the PNS
above roughly 10 times the neutrinospheric density was excised and
replaced by a gravitating point mass at the coordinate center
and an inner grid boundary, whose retraction from 60--80\,km shortly
after bounce to 20--30\,km after about 1\,s
was prescribed to mimic the shrinking of the PNS due to energy and lepton
number losses by neutrino emission.

Explosions of the models with chosen energies were initiated
by picking suitable values of the (time-dependent) neutrino luminosities
imposed spherically symmetrically
at the inner grid boundary. On the computational grid the transport
and matter interactions of neutrinos of all flavors
were approximated by the grey (non-spectral) transport
scheme described in Scheck et al.~\cite{Scheck:2006}, solving the
1D (spherically symmetric) transport equation in each angular grid
bin independently. This ``ray-by-ray'' approximation allowed for
taking into account directional variations of the neutrino fluxes.

With this setup the neutrino-driven explosions could be followed
until well beyond one second after bounce, at which time the
explosion energy was essentially saturated.
Starting from initial random seed perturbations (imposed
as cell-to-cell variations on the level of 0.1\% of the radial
velocity), large-scale asymmetries in the accretion flow between
the stalled shock and PNS ``surface'' developed (due to the action of
convective instability and the SASI) and supported the
onset of the explosion. Asymmetric accretion onto the nascent
neutron star (NS) continued considerably longer.
During the accretion phase a
sizable part of the $\nu_e$ and $\bar\nu_e$ emission (around
50\%) was produced in the hot, settling accretion layer between
the inner grid boundary and the gain radius, which bounds the
neutrino-heating layer from below.

\begin{figure}
\includegraphics[width=0.97\linewidth]{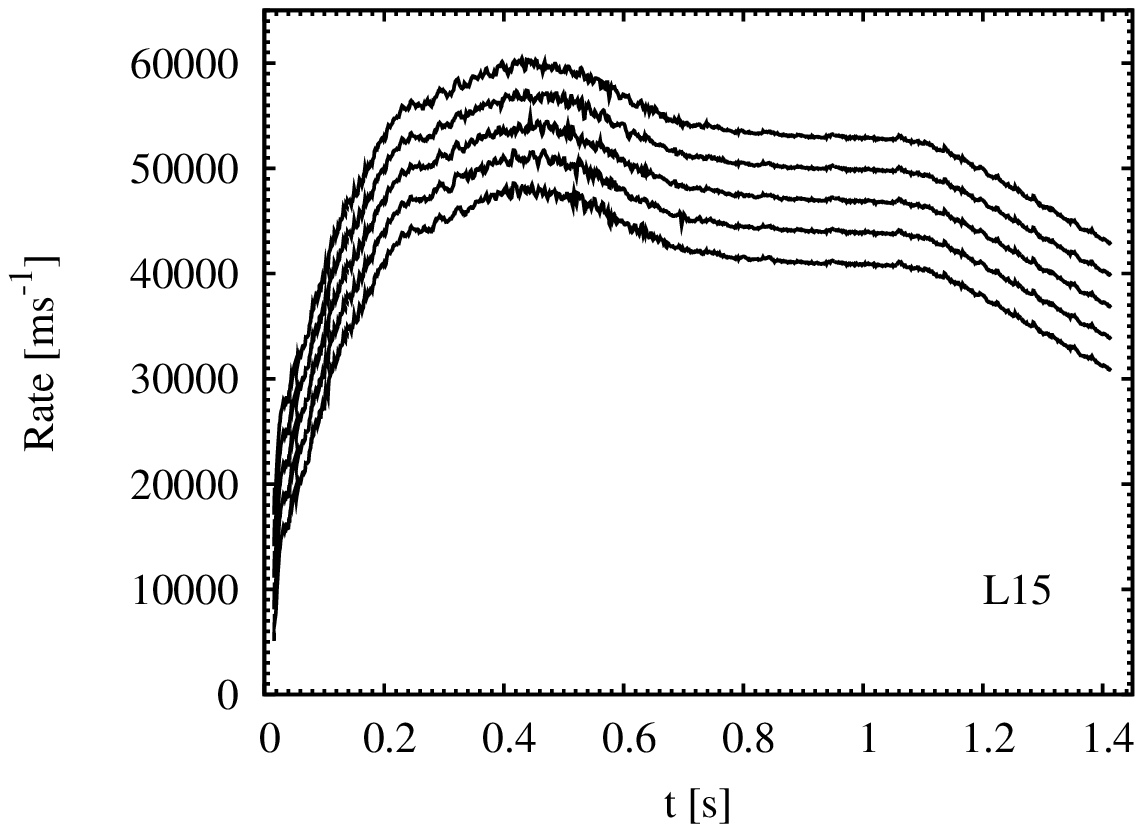}
\includegraphics[width=0.97\linewidth]{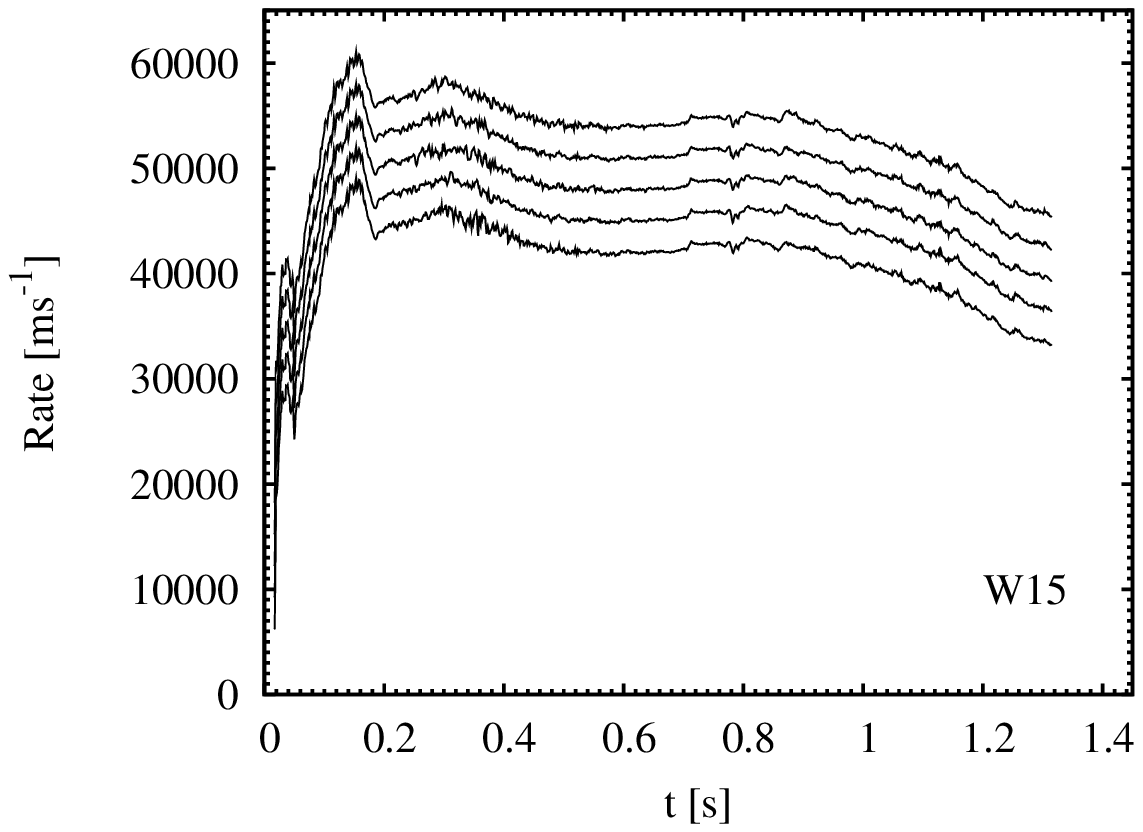}
\includegraphics[width=0.97\linewidth]{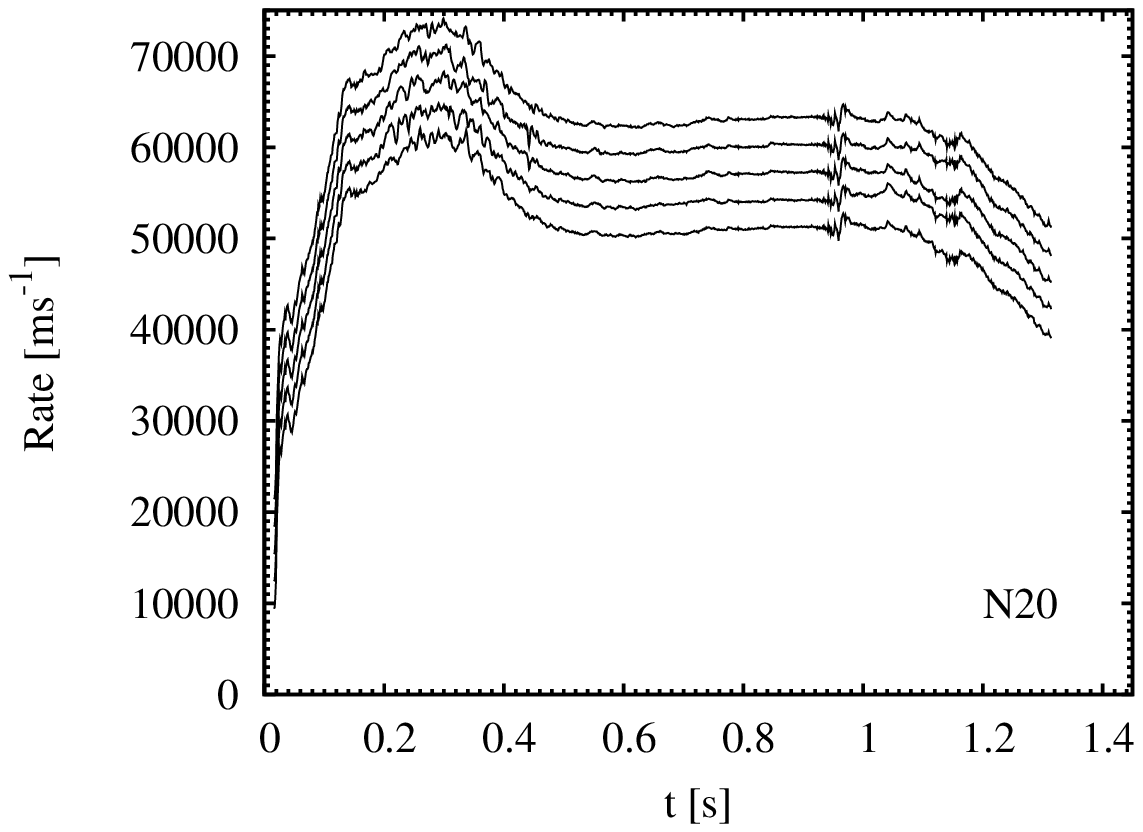}
\caption{\label{fig:all_rates} Event rates in
IceCube from our 3D models for a SN at 1~kpc.
The models are L15, W15 and N20 from top to bottom as indicated.
The rates are hemispheric flux integrals
for the five viewing directions,
with hemisphere 5 on top and other curves offset
consecutively by $-3000$ units.}
\end{figure}

The direction-dependent neutrino transport results were used
to deduce time-dependent information on the observable neutrino
luminosities for our models as described in Sect.~3
of~\cite{Mueller:2012jw}. To this end the emission
was integrated over the hemisphere seen by an observer from
a certain direction, approximately accounting for the limb
darkening effect that is associated with a non-isotropic
angular distribution of the neutrinos in the layer of
neutrino-matter decoupling. For the analysis in this paper we
chose five viewing directions, which are orthogonal to each other
and can be considered as representative, because the neutrino
emission in the (nonrotating) 3D models does not
possess any preferred direction. This can be verified from
Fig.~\ref{fig:all_rates}, where the expected IceCube detection rates
show similar time dependence and differ only in details for
all investigated viewing angles. In addition to the
observable luminosities the grey neutrino
transport also provides the mean energies of the radiated neutrinos
as the ratio of neutrino energy flux to neutrino number flux.

\subsection{Neutrino emission from SN cores}   \label{sec:SNneutrinos}

The count rates displayed in Fig.~\ref{fig:all_rates} reflect the
familiar behavior of the $\bar\nu_e$ emission from a SN core. While
the mean neutrino energies increase with time mostly monotonically
after bounce, the luminosities exhibit a steep rise over a timescale
of typically 100\,ms, followed by a broad accretion hump of
400--600\,ms, whose length depends on the duration of the postbounce
accretion phase, and finally, after the onset of the explosion, the
luminosities make a transition to a continuous decline, which marks
the beginning of the PNS cooling evolution (see also Fig.~9 in
\cite{Mueller:2012jw}). Besides rapid fluctuations on timescales
between several and tens of ms and maximum amplitudes on the 5\%
level, the event rates also show modulations on longer timescales.
These are connected to variations of the mass infall rate to the
shock, which is determined by the density structure of the
collapsing progenitor core. In particular the W15 model reveals the
effects of sharp composition-shell interfaces. When the latter fall
through the shock, the mass accretion rate drops rapidly and the
accretion luminosity follows this behavior with a short delay (the
most prominent of these features is seen in the W15 model between
180 and 200\,ms). At late times ($t \gtrsim 800$--1000\,ms)
quasiperiodic, low-level variability is associated with convective
activity close to the neutrinospheric layer inside the PNS, and
aperiodic, stronger outbursts are a consequence of episodic,
asymmetric fallback of gas to the PNS.

The fast variability with frequencies between $\sim$30\,Hz and
$\sim$200\,Hz during the phase of massive accretion is caused by
large-scale, intermittent accretion funnels, which carry infalling
matter from the SN shock front to the newly formed NS. These
asymmetrically distributed, strongly time-dependent accretion
streams are associated with the convective overturn in the
neutrino-heating layer and can be influenced by SASI motions of the
shock and postshock mass distribution. The corresponding activity
between PNS and shock becomes particularly strong after 100--200\,ms
of postbounce evolution in all investigated models. At this time
neutrino-flux variations develop on different angular scales, but
soon the emission asymmetry becomes dominated by low spherical
harmonics modes ($\ell = 1$--4). For hundreds of milliseconds, well
beyond the onset of the explosions, the neutrino emission exhibits
strong intensity maxima where a few large, well localized accretion
downdrafts impact the neutrinospheric layer (for details, see
\cite{Mueller:2012jw}). Because the location of these hot spots can
be relatively stationary (especially when the shock expansion has
set in and the convective bubble pattern begins to freeze out), the
luminosity variations visible for an observer from a certain
direction carry information about the typical coherence timescale of
the accretion flows feeding the asymmetric emission. This coherence
timescale is roughly set by the infall time of matter through the
accretion funnels, $\sim$$(R_\mathrm{s}-R_\mathrm{NS})/|v_r|$, and
is estimated to be tens of milliseconds for representative values of
the shock radius, $R_\mathrm{s}$, PNS radius, $R_\mathrm{NS}$, and
radial velocity, $v_r$.

\subsection{Neutrino signal in IceCube}           \label{sec:rate_calc}

As in our previous study~\cite{Lund:2010}, we consider the neutrino
signal in IceCube because it produces the largest event statistics
of all operating SN neutrino detectors. The grey neutrino transport
approximation employed in our models only provides average neutrino
energies without detailed spectral information and we assume thermal
distributions. Following our previous estimate, the expected signal
rate then is
\begin{equation}\label{eq:eventrate}
R_{\bar\nu_e}=153~{\rm ms}^{-1}\,
\frac{L_{\bar\nu_e}}{10^{52}~{\rm erg}~{\rm s}^{-1}}
\left(\frac{10~{\rm kpc}}{D}\right)^2
\left(\frac{E_{\rm av}}{12~{\rm MeV}}\right)^2\,.
\end{equation}
Even without a SN signal, the detector produces a background (bkgd)
event rate of approximately
\begin{equation}
R_{\rm bkgd}=1340~{\rm ms}^{-1}\,.
\end{equation}
It must be added to the SN induced signal to obtain the total event
rate $R_{\rm tot}$. We display the expected counting rates for an
assumed distance of 1~kpc in Fig.~\ref{fig:all_rates} for the five
observation directions mentioned earlier.

Numerical models in 2D showed strong dipolar and quadrupolar
motions~\cite{Marek:2008qi, Marek:2007gr} that imparted excursions
in the neutrino luminosity of up to 30\% and up to 15\% in the
mean neutrino energy.
The resulting event-rate fluctuations were almost up to
20\%. This meant that a detailed power spectrum could be obtained
for a SN out to a distance of 10 kpc \cite{Lund:2010}. In the case
of our 3D models, the SASI and convective flows are much less
coherent in space. Consequently, the amplitudes of variations in
the neutrino luminosities and mean
energies are considerably smaller. This reduction reflects in the
event rates shown in Fig.~\ref{fig:all_rates}, where the
fluctuations amount to only a few percent. We
therefore focus our analysis on a fiducial SN distance of 1~kpc
instead of the previously used 10~kpc.

The sharp dips in the event rate seen particularly for the W15 model
are caused by burning-shell interfaces falling through the accretion
shock. The corresponding drops in density reduce the accretion rate
and the neutrino emission. The effect of shell interfaces was
clearly visible in spherical simulations but appeared to be obscured
or dwarfed in 2D by the fairly large SASI-induced emission
fluctuations. In 3D, where the SASI effects are smaller, the shell
interfaces become visible once again.

In contrast to the axisymmetric 2D simulations,
there is no preferred spatial direction in 3D (if rotation is
absent in the stellar core), and the event rates in all viewing
directions are very similar (see Fig.~\ref{fig:all_rates}).
In the following we will always use ``hemisphere 5'' as our
benchmark case.

We focus our analysis on the accretion phase where the neutrino
signal and its variations are largest. On the other hand, convective
overturn and SASI activity take some time to develop. For an
analysis interval we therefore begin at 100~ms post bounce until
somewhat after the explosion takes off. Specifically, we will use
time intervals $\tau=300$~ms (W15), 350~ms (N20) and 500~ms (L15)
as listed in Table~\ref{tab:datachart}.

\section{Power spectra}\label{sec:power}

\begin{figure*}
\includegraphics[width=0.97\columnwidth]{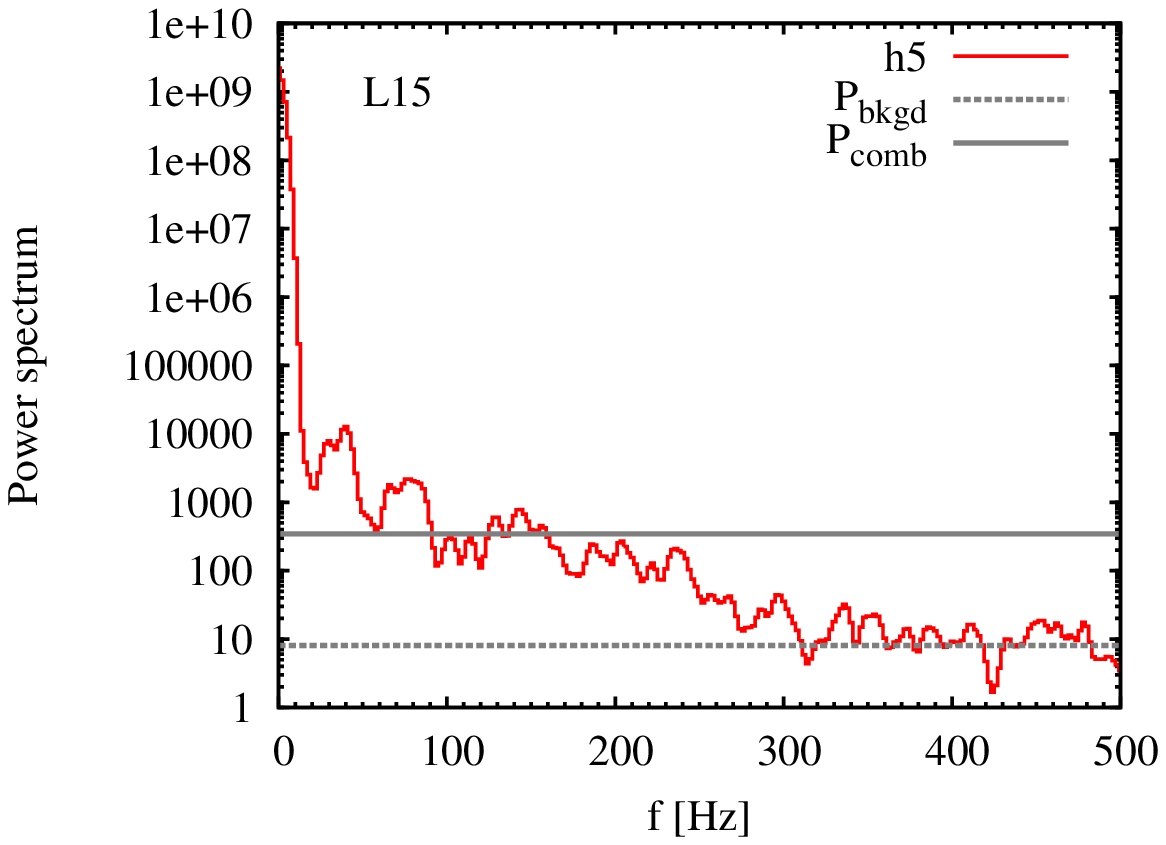}\hskip1cm\includegraphics[width=0.97\columnwidth]{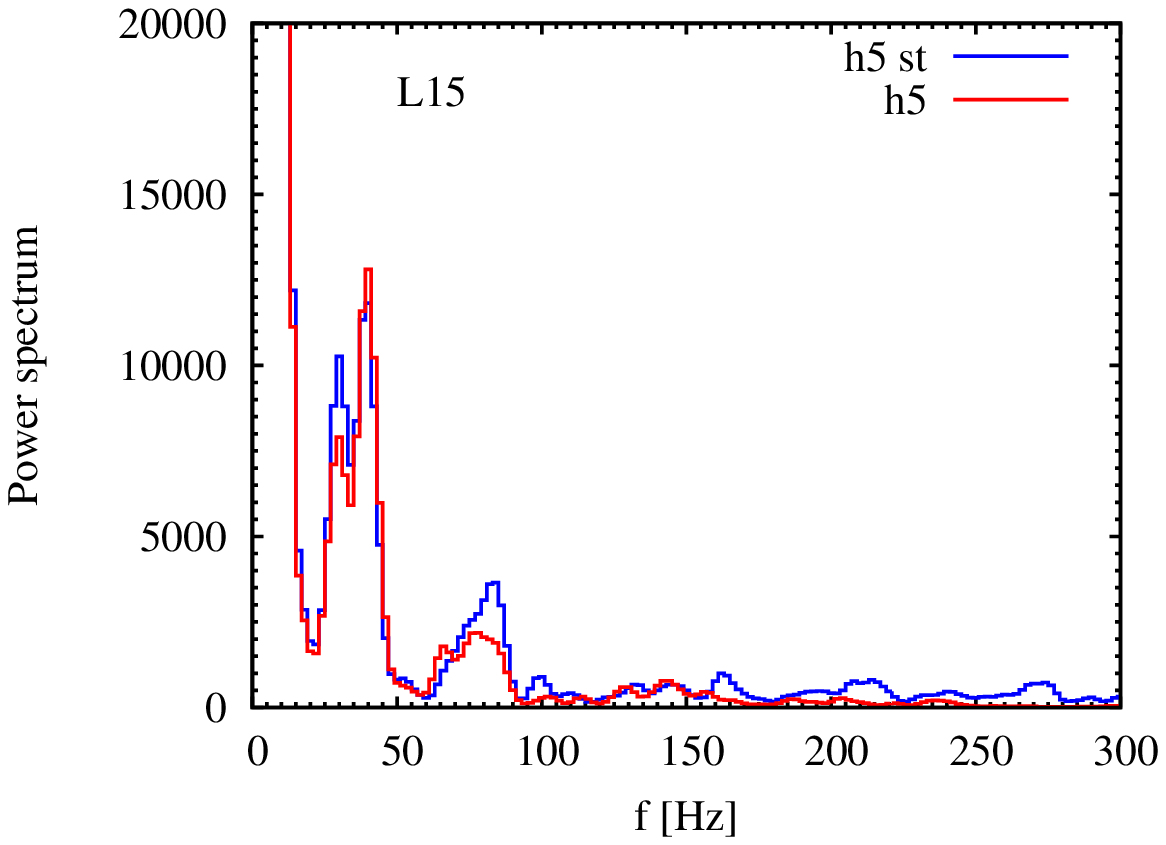}
\includegraphics[width=0.97\columnwidth]{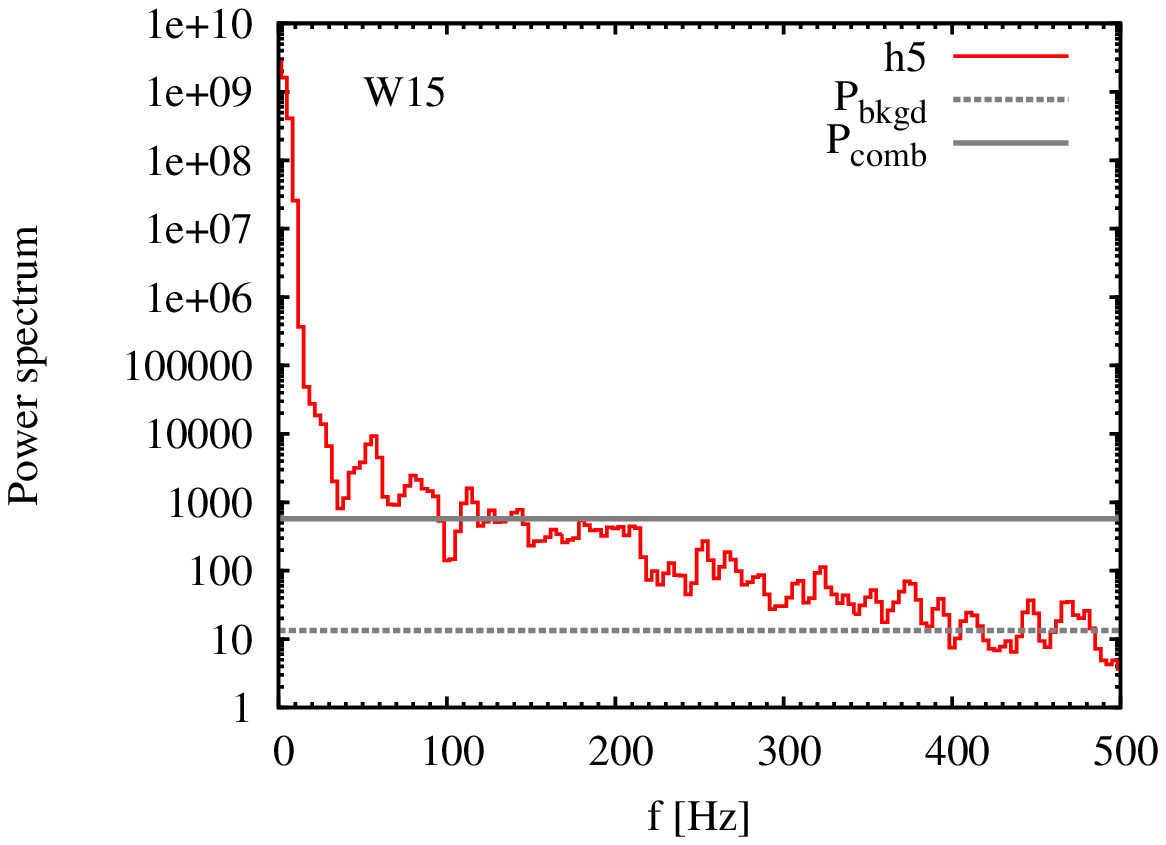}\hskip1cm\includegraphics[width=0.97\columnwidth]{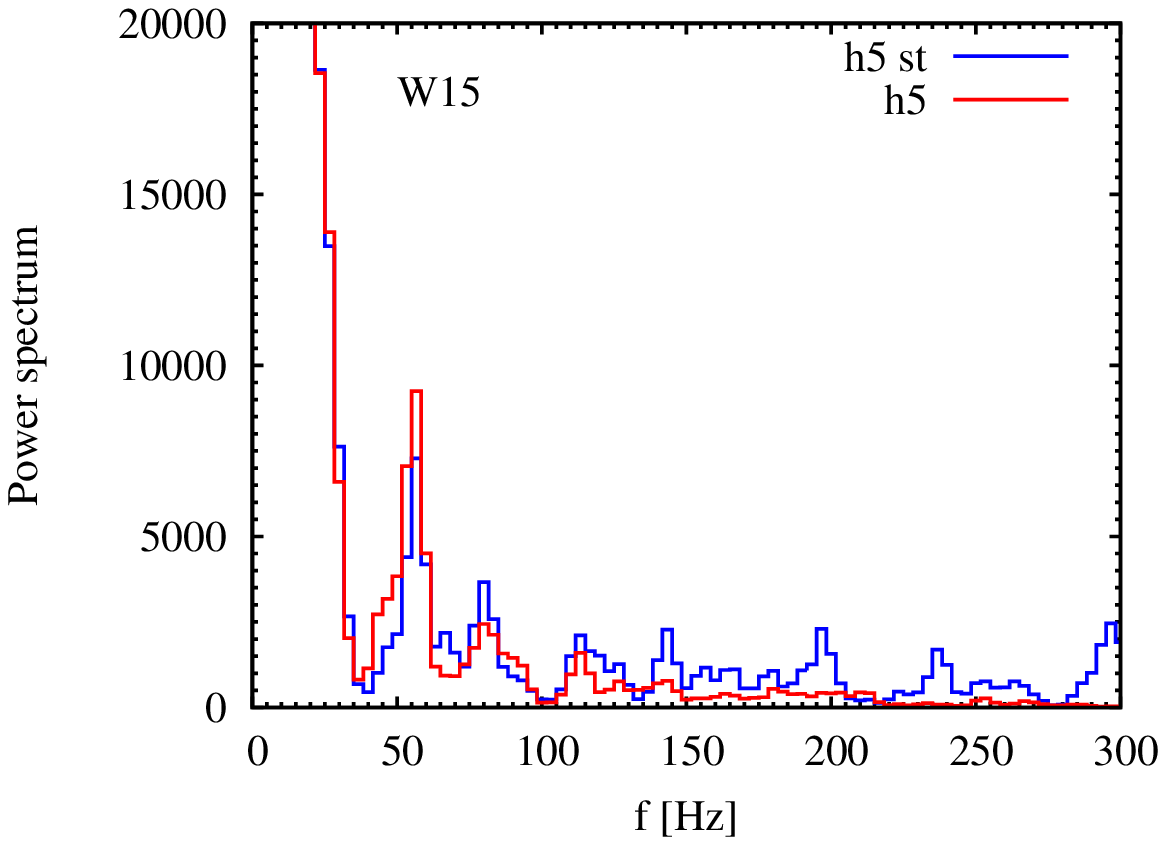}
\includegraphics[width=0.97\columnwidth]{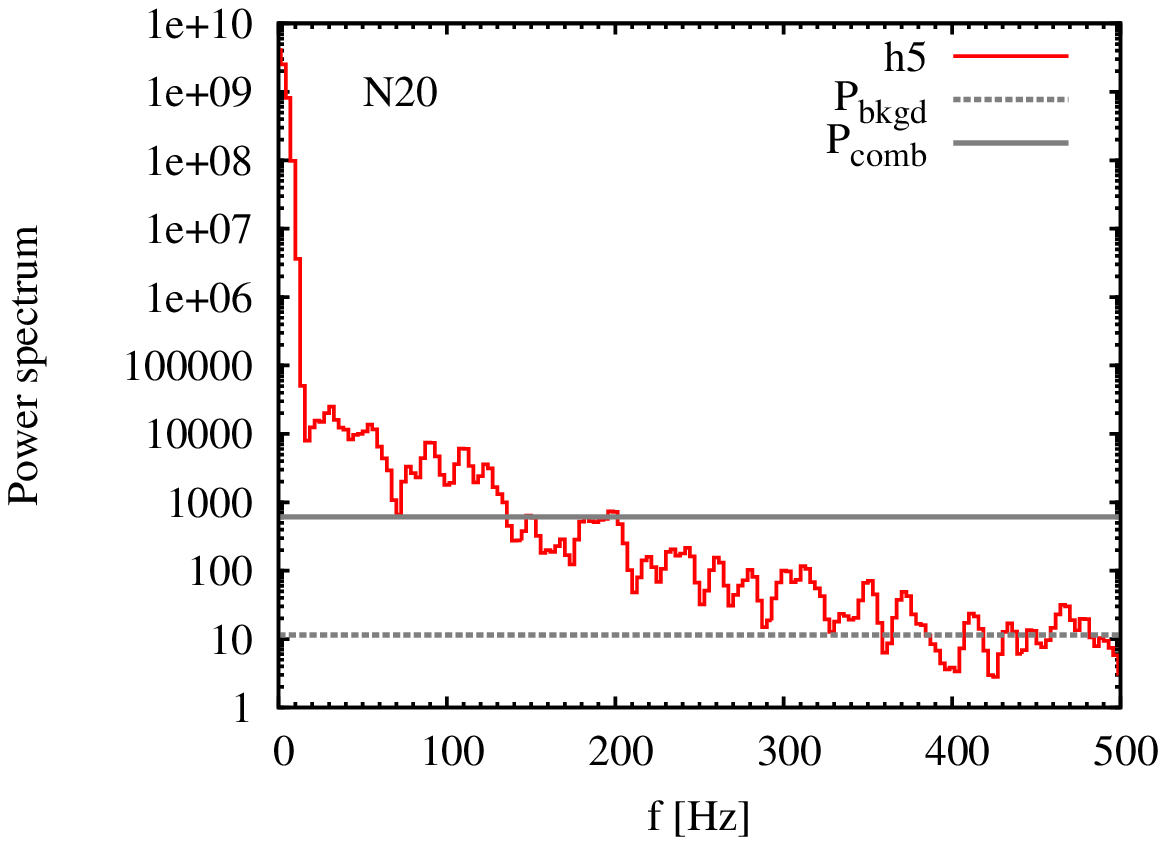}\hskip1cm\includegraphics[width=0.97\columnwidth]{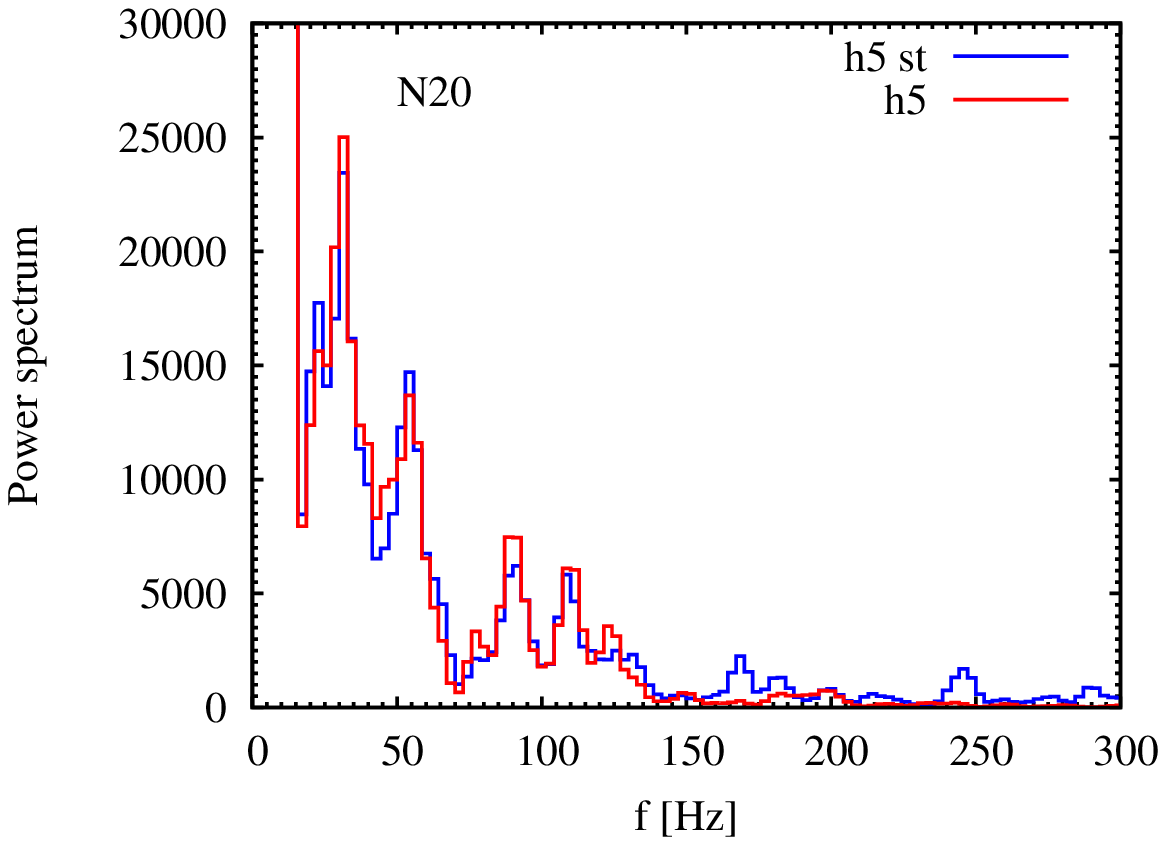}
\caption{\label{fig:power} (color online). Power spectra for our three models as
indicated, in each case for hemisphere 5 and a distance of 1~kpc. We
also show the average power levels for the shot noise arising from the
IceCube dark current ($P_{\rm bkgd}$, dashed gray line) and from
the average signal plus background fluctuations ($P_{\rm comb}$,
solid gray line). The right panels use a
linear scale and in addition show sample realizations (blue) where
shot noise has been included.}
\end{figure*}

We show the power spectra of our three benchmark cases in
Fig.~\ref{fig:power}, i.e.\ the ``hemisphere 5'' emission of the
three models. Each power spectrum has been smoothed in the frequency
domain by a running Gaussian ($\sigma=2.5$~Hz) to enhance the
visibility of the overall features.

We recall our definition of the Fourier transform of a signal rate
$R(t)$ that is sampled in $N_{\rm bins}$ bins of equal width
$\Delta=\tau/N_{\rm bins}$,
\begin{equation}
h(f_k) = \Delta \sum_{j=0}^{N_{\rm bins}-1} R(t_j)\,\E^{\I 2 \pi t_j k \delta f}\,.
\end{equation}
The frequencies are $f_k=k/\tau=k\delta f$ with $k=0,\ldots,N_f$ and
$N_f=f_{\rm max}/\delta f$. Here $f_{\rm max}=1/2\Delta$ is the
Nyquist frequency. Since $\delta f=1/\tau$ we have $N_f=N_{\rm
bins}/2$. We give the values for the frequency resolution $\delta f$
for our three cases in Table~\ref{tab:datachart}. We use
$\Delta=1$~ms, providing a Nyquist frequency of $f_{\rm max}=500$~Hz
as an upper cutoff\footnote{Our chosen 1~ms binning is more convenient 
than IceCube's native 1.64~ms bin width. The only change is a slightly
increased Nyquist frequency which is inconsequential for our 
discussion}. The spectral power, a dimensionless number, is
\begin{equation}
P(f_k)=2\,\frac{|h(f_k)|^2}{N_{\rm bins}^2}
\end{equation}
except for $f=0$ and $f_{\rm max}$ where the factor of 2 drops out.

We use a Hann window function to suppress edge effects. Note that
our power spectrum is in absolute units, not relative to the average
signal. The Fourier amplitude at $f=0$ is the average rate per bin:
$N_{\rm events}/N_{\rm bins}$. The power at $f=0$ is therefore
$P_0=(N_{\rm events}/N_{\rm bins})^2$, which fixes the overall
normalization and is given for our three cases in
Table~\ref{tab:datachart}.

Returning to Fig.~\ref{fig:power}, the logarithmic panels show the
familiar fall off at high frequencies previously seen in Figs.~5 and
6 of Ref.~\cite{Lund:2010}. This decrease is caused by the spatial
integration over the entire visible hemisphere of the ``boiling''
stellar matter. The linear panels show significant peaks at familiar
frequency values. In the L15 model, we see a wide double peak at
25--50~Hz and a broad bump stretching roughly from 60 to 85~Hz, as
well as a few peaks at higher frequencies. Model W15 displays a
clear peak at 55~Hz, along with peaks at 80~Hz and $\sim$112~Hz. A
significant peak around 55~Hz is also visible in N20, but this time
it is accompanied by an even larger peak at 30~Hz. Further strong
peaks are visible at 90 and 110~Hz, and two weaker ones at 75 and
125~Hz.

The frequencies of the peaks in the power spectra are similar to
those that we found in our previous analysis of 2D
models~\cite{Lund:2010}. This is remarkable because the dynamics of
the postshock layer and the morphology of the accretion flow in the
3D simulations appear to be considerably different from those
observed in 2D calculations. While the 2D dynamics is strongly
influenced by violent, large-amplitude SASI sloshing motions along
the symmetry axis, such a phenomenon is not visible in the 3D runs.
The much larger amplitudes of the neutrino-emission variations in 2D
imply that the SASI sloshing motions are able to enhance the
luminosity and energy fluctuations. A strong effect is fostered by
the large spatial coherence scale of the 2D flow pattern, which
channels accreted matter alternately toward the north and south
polar regions~\cite{Marek:2008qi}. The similarity of the peak
frequencies, however, suggests that in 2D as well as in 3D the
modulation of the hot spots of neutrino emission is governed by the
same basic effect, namely the temporal coherence of the asymmetric
accretion pattern, which is linked to the timescale of mass
advection from the shock to the PNS surface (see
Sect.~\ref{sec:SNneutrinos}). For similar shock and NS radii and
similar advection velocities, the emission variability occurs on
comparable characteristic timescales. In this context the slightly
lower frequency of the first peak in models L15 and N20 (around
30~Hz) compared to model W15 (near 50\,Hz) can be explained by the
smaller shock stagnation radius in the W15
case~\cite{Mueller:2012jw}.

Although these arguments appear plausible, a satisfactory
theoretical understanding of the peaks in the power spectra
is still lacking. The sequence of peaks might point to the
existence of a fundamental frequency and higher harmonics,
but the underlying reason is unclear.
Moreover, a detailed comparison of dimension-dependent
effects will require 2D and 3D modeling with the same
level of sophistication instead of the approximations used in
the present set of 3D runs as described in Sect.~\ref{sec:models}.

\section{Detectability}\label{sec:shotnoise}

\begin{figure}
\includegraphics[width=0.97\linewidth]{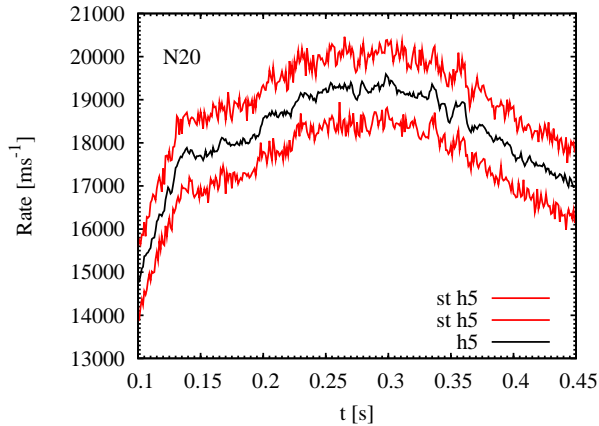}
\caption{\label{fig:noise_time} (color online). Event rate for a SN at 2~kpc, including
IceCube dark current. The black line is the signal for the N20 model,
the red lines are two realizations when shot noise is added
(curves are offset by $\pm$800 units).}
\end{figure}

The detectability of features in the power spectrum depends on their
significance relative to random fluctuations (shot noise). In
Fig.~\ref{fig:noise_time} we show an example of a SN signal at 2~kpc
without shot noise (black line) and two realizations with shot noise
added, each time using our usual 1~ms bins. Visually the shot noise
competes with the intrinsic signal variations and indeed a distance
of 2~kpc is about the limiting case where signal variations can
still be detected.
The probability of a future galactic core collapse SN being within 
2~kpc of Earth is only a few percent of the total galactic 
rate \cite{Mirizzi:2006xx} of a few per century. Therefore, the prior 
probability of such an event is, of course, very small, yet not 
completely excluded.

As explained in our previous paper~\cite{Lund:2010}, the average of the
shot noise provides a flat power spectrum. In the framework of
our definitions, and using a Hann window function, the power-level
of the shot noise is
\begin{equation}\label{eq:shotnoise}
P_{\rm shot}=
3N_{\rm events}/N_{\rm bins}^2\,.
\end{equation}
Here $N_{\rm events}$ is the total number of events in the analysis
interval of duration $\tau$.

One unavoidable source of shot noise is the IceCube dark current
with the rate $R_{\rm bkgd}$, providing the shot noise level $P_{\rm
bkgd}$ shown for our three cases in Table~\ref{tab:datachart}. The
signal variations in our models are now so weak that we need a
relatively close SN to detect these variations, which in turn means
that the SN signal contributes much more to the counting rate than
the dark current. Therefore, the dominant source of shot noise is
now the signal itself and we give the corresponding power levels in
Table~\ref{tab:datachart} as $P_{\rm comb}$ which includes the
combined shot noise from the signal itself and the dark current. We
also show the average power levels $P_{\rm bkgd}$ and $P_{\rm comb}$
in the left panels of Fig.~\ref{fig:power}. It is immediately
obvious that even at 1~kpc one cannot hope to detect significant
signal power at frequencies exceeding some 150~Hz.

The same point can be made by the right panels of
Fig.~\ref{fig:power} where we show the power spectra on a linear
scale, together with typical realizations including shot noise for a
SN at 1~kpc. For the signal with added shot noise the power spectrum
clearly retains the intrinsic low frequency peaks. In addition,
relatively strong peaks appear for frequencies $f\gtrsim 150$~Hz.

\begin{figure}[t!]
\includegraphics[width=0.97\linewidth]{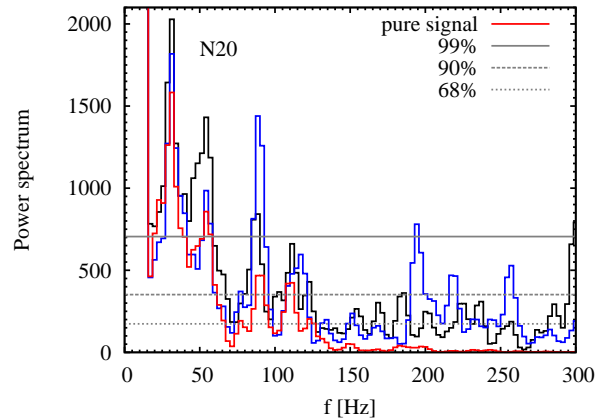}
\caption{\label{fig:exclusion_levels} (color online). Power spectrum for the signal
from hemisphere 5 of the N20 model,
for a SN at 2~kpc (red line) and two realizations with added
statistical noise (blue and black lines). The 99\%, 90\% and 68\%
levels are also shown (grey lines).}
\end{figure}

To evaluate how likely it is that a peak caused by fluctuations
mimics a real peak, we calculate the power level where 99\% of
random peaks should lie below. From the expressions given in our
previous paper we find these limiting power levels $P_{99}$ and give
them in Table~\ref{tab:datachart}. In
Fig.~\ref{fig:exclusion_levels} we show a pure signal (red line) and
two signals with addition of different shot noise realizations
(black and blue curves). The grey lines give the levels where 99\%,
90\% and 68\% of the purely random peaks should lie below. We see
that the main peaks at $f\sim$30~Hz, 50~Hz, and 90~Hz can all be
identified.

The power of the shot noise relative to the signal scales linearly
with distance, i.e.\ at twice the distance, the shot noise is twice
larger relative to the signal power. We conclude that at distances
exceeding 2~kpc it will be nearly impossible to distinguish
intrinsic signal variations from random fluctuations caused by the
limited counting rate.

\section{Summary and conclusions}\label{sec:conclusion}

We have shown that also for 3D SN models time variations of the
neutrino emission lead to distinct peaks in the power spectra
as previously seen for 2D results~\cite{Lund:2010}.
Despite considerably smaller amplitudes of the luminosity and mean
energy fluctuations than in the 2D simulations, these features will
still be detectable in the IceCube event rate for a future galactic
SN within 2\,kpc distance. Up to frequencies of about 150\,Hz such
features will hint to intrinsic flux variations, whereas above
this frequency they will, with a high probability, be associated
to statistical effects in the measurement of the signal.

A thorough discussion is hampered by the simplifications made in the
3D models and the small set of investigated cases. In particular,
the neutrino signals evaluated in this work are based on an
approximate grey treatment of the neutrino transport in models, in
which the high-density core of the PNS was excised and replaced by a
boundary condition. Nevertheless, the signals exhibit all familiar
features also predicted by more sophisticated transport schemes in
fully self-consistent 1D and 2D simulations. The largest emission
variations are observed during the postbounce accretion phase until
shortly after shock revival, when the SN core is stirred by violent
hydrodynamical instabilities. Large-scale asymmetries and
intermittent behavior characterize the accretion flows to the
nascent NS during this phase. Interestingly, the power spectra of
the 3D models show similar peaks in the same frequency range as
those found in 2D, although the flow dynamics in 3D appears to be
largely different. In particular, the strong SASI sloshing motions
that are constrained to the symmetry axis in the 2D case and that
are responsible for big emission variations, cannot be seen
prominently in the 3D simulations.

If the existence of the frequency peaks is consolidated by better 3D
models, these features will offer a powerful diagnostic for the
dynamics in the SN core and the explosion mechanism. The role of
turbulence and violent nonradial mass motions in the postshock layer
during the accretion phase and on the way to shock revival is
presently a matter of vivid debate
\cite{Burrows:2012,Murphy:2012db,Mueller:2012jh}. A better
understanding of the characteristic imprints of the SASI and of
convective overturn motions on the neutrino emission in 3D is
therefore desirable.

Our previously investigated 2D models together with the 3D
simulations analyzed in this work can be considered to span the
range between optimistic and pessimistic cases with respect to the
possible level of signal variability. It is not excluded that fully
self-consistent 3D models will reveal higher fluctuation amplitudes
of the neutrino emission. Stronger SASI activity and especially the
development of spiral modes \cite{Blondin:2007nat}, whose growth
could be fostered by even a relatively small amount of rotation in
the SN core, can not be excluded on grounds of the small set of
presently available 3D simulations.

\section*{Acknowledgments}

We thank the Institute for Nuclear Theory at the University of
Washington for its hospitality and the Department of Energy for
partial support while this work was completed during the INT Program
INT 12-2a ``Core-Collapse Supernovae: Models and Observable
Signals''. In Munich and Garching, we acknowledge partial support by
the Deutsche Forschungsgemeinschaft under grant TR-7 ``Gravitational
Wave Astronomy'' and the Cluster of Excellence EXC-153 ``Origin and
Structure of the Universe.'' T.L.\ acknowledges support from the MPA
in Garching and the MPP in Munich, as well as hospitality at Aarhus
University, while much of this work was done.


\end{document}